\documentclass[aps,pra,amsmath,twocolumn,amssymb,titlepage]{revtex4-1}

\usepackage{graphicx}
\usepackage{color}
\usepackage{graphics}
\usepackage{epsfig}
\usepackage{subfigure}

\usepackage{epstopdf}
\usepackage{amsfonts}

\usepackage{amssymb}
\usepackage{amsmath}

\usepackage{subfigure}
\usepackage{multirow}
\usepackage{tabularx}
\usepackage{array}
\usepackage{ulem}
\usepackage{bm}
\usepackage{hyperref}

\providecommand{\U}[1]{\protect\rule{.1in}{.1in}}

\begin{document}
	\title{Subnanosecond magnetization reversal of magnetic nanoparticle driven by chirp microwave field pulse}
	\author{M. T. Islam, X. S. Wang, Y. Zhang}
	\author{X. R. Wang}
\email{[Corresponding author:]phxwan@ust.hk}
	\affiliation{Physics Department, The Hong Kong University of
		Science and Technology, Clear Water Bay, Kowloon, Hong Kong}
\affiliation{HKUST Shenzhen Research Institute, Shenzhen 518057, China}
		
\begin{abstract}
We investigate the magnetization reversal of single-domain magnetic
nanoparticle driven by linear down-chirp microwave magnetic field pulse.
Numerical simulations based on the Landau-Lifshitz-Gilbert equation
reveal that solely down-chirp pulse is capable of inducing subnanosecond
magnetization reversal. With certain range of initial frequency and chirp
rate, the required field amplitude  is much smaller than that of
constant-frequency microwave field. The fast reversal is because the 
down-chirp microwave field acts as an energy source and sink for the magnetic 
particle before and after crossing over the energy barrier, respectively.
Applying a spin-polarized current additively to the system further
reduces the microwave field amplitude. Our findings provide a new way to
realize low-cost and fast magnetization reversal.
\end{abstract}

\maketitle

\section{ INTRODUCTION}
Magnetization reversal of single-domain magnetic nanoparticle draws
significant attentions because of its application in high-density data
storage \cite{S2000,SI2001,D2002} and processing \cite{B2001}.
Fast reversal of single-domain magnetization with minimal energy cost
is the ultimate demand in device applications. To achieve high thermal
stability and low error rate, high anisotropy materials are used
so that magnetic nanoparticles have high energy barrier \cite{nature}.
It is difficult but essential to find out how to achieve fastest
magnetization reversal for high-anisotropy magnetic nanoparticles with
energy cost as low as possible. Over the last few years, a number of
theoretical schemes have been proposed and some of them have been verified
by experiment. In the early years, constant magnetic field was used
as driving force to reverse the magnetization \cite{ 1book, XR2005}, 
but the reversal time is too long \cite{1book} and it suffers from 
scalability problems because the energy consumption per unit area 
increases as the device feature size decreases. Since the the discovery of 
spin transfer torque (STT) \cite{STT}, people prefer to deploy spin polarized 
electric current \cite{M1998,J1999,Z2004,A2000, X2000, E1999, JA2000, R2004} 
to reverse the magnetization, and devices based on STT magnetization
reversal have been fabricated. However, large current density is required
for fast reversal so that significant Joule heat shall limit the device
durability and reliability of the device \cite{HMORISE,Tshinjo,Joule}.
If the direction of the magnetic field or current varies with time in
a designed way, the field/current amplitude or switching time can be
much lower \cite{XR2007} than that of constant field/current.
But it is strenuous to generate such kind of fields/currents in practice.
Microwave magnetic field, either with or without a polarized electric
current, is another controlling knob for magnetization reversal
\cite{Bertotti,IEEE,XR2006}. A microwave of constant frequency itself
can reverse a magnetization through synchronization \cite{XR2005}.
Large field amplitude is required and the reversal process is relatively
slow \cite{T2016,C2003,SI2006,SO2008,TT2013}. Recently, there are
several theoretical approaches demonstrating the magnetization reversal
by microwaves of time-dependent frequency \cite{KR,ZM, Suto17,GK, NBarros,Lcai,
Topical view}. However, the existing strategies are either too
complicated or lack of clear physical pictures.
In this paper, we show that a circularly polarized down-chirp microwave
pulse (a microwave pulse whose frequency decreases with time) can
effectively reverse a magnetization. For a nanoparticle of a
high uniaxial anisotropy (coercive field $h_k\sim0.75$ T), subnanosecond
magnetization reversal can be achieved. With proper choice of initial
frequency and chirp rate, the microwave field amplitude required for
subnanosecond magnetization reversal is only several tens of mT, much
smaller than that required for a constant-frequency microwave field.
Because through the reversal process, the anisotropy field decreases as
the component of magnetization along the easy axis decreases, the
intrinsic precessing frequency of the magnetization also decreases.
After passing the equator, the intrinsic precessing frequency changes
its sign. The down-chirp microwave field acts as an energy source (sink)
before (after) the magnetization reaches the equator in the fast
magnetization reversal. We further show that linear-polarized down-chirp
microwave field pulse is also capable of reversing a magnetization fast.
We also demonstrate a spin-polarized current can work together with the
down-chirp microwave field pulse so that both applied current density
and microwave amplitude are low enough.

\section{\textbf{Model and Methods}}	

We consider a spin valve with free and fixed ferromagnetic layers and a 
nonmagnetic spacer in between, as shown schematically in Fig. 1(a). 
Both fixed and free layers are perpendicularly magnetized. 
The magnetization direction of the fixed layer $\mathbf{p}$ is pinned to
be upward, $\mathbf{p}=\hat{\mathbf{z}}$ ($\hat{\mathbf{z}}$ is the unit
vector along $z$ direction). The magnetization of free layer is treated 
as a macrospin with magnetization direction $\mathbf{m}$ and saturation 
magnetization $M_\text{s}$. The macrospin approximation is valid 
for device size smaller than 100 nm \cite{Yzhang}. 
The Landau-Lifshitz-Gilbert (LLG) equation governs the magnetization 
dynamics in the free layer in the presence of spin polarized 
current and microwave magnetic field \cite{XR2005,XR2006,XR2007,T2016}
\begin{equation}
\dfrac {d\mathbf{m}}{dt}=-\gamma \mathbf{m}\times \mathbf{H}_{\text{eff}}-\gamma \text{h}_{\text{s}} \mathbf{m}\times (\mathbf{p}\times\mathbf{m}) +\alpha \mathbf{m}\times \dfrac {d\mathbf{m}}{dt},
\label{llg}
\end{equation}
where $\gamma$ is the gyromagnetic ratio, $\alpha$ is the Gilbert damping constant.
The total effective field \(\mathbf{H}_{\text{eff}}\) consists of the microwave 
magnetic field \(\mathbf{H}_{\text{mw}}\) and the anisotropy field \(\mathbf{H}_{\text{K}}=H_\text{K}m_z\hat{\mathbf{z}}\), i.e., \(\mathbf{H}_{\text{eff}}=\mathbf{H}_{\text{mw}}+\mathbf{H}_{\text{K}}\) .
\(h_{\text{s}}\) represents the intensity of spin transfer torque (STT) \cite{STT},
 \begin{equation}
 	h_{\text{s}}=\dfrac {\hbar P j}{2 e \mu_0M_{\text{s}} d}\label{cur},
 \end{equation}
where $j$, \(e\), \(P\), and \(d\) represent the current density,
electron charge, spin polarization of current, and thickness
of the free layer, respectively. $\hbar$ is the Plank's constant and
$\mu_0$ is the vacuum permeability. In the following study, the 
parameters are chosen from typical experiments
on the microwave driven magnetization reversal as $M_\text{s}=10^{6}$
\(\text{A}/\text{m}\), \(H_{\text{k}}=0.75\) T,
\(\gamma = 1.76\times10^{11}\) rad/(T\(\cdot\)s), \(P = 0.6\),
\(\alpha=0.01\), \(d = 2\) nm.

The microwave field $\mathbf{H}_\mathbf{mw}$ and
the spin transfer torque are non-conservative forces. They do work on the macrospin.
We first consider solely microwave-driven magnetization reversal. 
Without STT term, the rate of energy change of the macrospin is expressed as
\begin{equation}
\dot{\varepsilon}=-\dfrac {\alpha}{1+\alpha^{2}}|\mathbf{m}\times \mathbf{H}_{\text{eff}}|^{2}-\mathbf{m}\cdot\dot{\mathbf{H}}_{\text{mw}}.
\label{eng1}
\end{equation}	
The first term is always negative because of the positive damping factor whereas 
the second term can be either positive or negative for a time-dependent field. 
In other words, the microwave field can be either an energy source or an energy 
sink, depending on the relative angle between the instantaneous magnetization 
direction and the time derivative of the microwave field.

Due to the easy-axis anisotropy, the magnetization has two stable equilibrium
states, $\mathbf{m}=\pm\hat{\mathbf{z}}$, corresponding to two energy minima. 
The magnetization reversal is to reverse the magnetization from
one equilibrium state to the other. Thus, the magnetization
has to be driven to overcome an energy maximum at the equator $m_z=0$. 
Before $\mathbf{m}$ reaches the equator, it gains energy from external forces.
After $\mathbf{m}$ passes the equator, it releases energy through damping or
through the negative work done by external forces. For a microwave field, the ideal 
case for fast magnetization reversal is that the microwave always synchronize 
to the magnetization motion so that $\mathbf{m}\cdot\dot {\mathbf{H}}_{\text{mw}}$ 
keeps maximal before reaching the equator and keeps minimal 
after passing the equator. However, this is difficult to achieve in practice. 
We notice that the internal effective field due to anisotropy
is $\mathbf{H}_{\text{K}}=H_{\text{K}}m_z\hat{\mathbf{z}}$, which corresponds
to a resonant frequency proportional to $m_z$. During a magnetization reversal
from $m_z=1$ to $m_z=-1$, the resonant frequency first decreases, then reaches
0 at the equator, then increases again with opposite precession direction.
This leads us to consider a down-chirp microwave pulse, whose frequency decreases
with time. If the rate of frequency change matches the magnetization precession,
the microwave field roughly accommodates the magnetization precession,
and acts as an energy source (sink) before (after) reaching the equator,
so that the magnetization reversal can be fast.

In order to demonstrate the feasibility of above scenario, we apply a
circular-polarized down-chirp microwave pulse on the system and 
numerically solve the LLG equation using MuMax3 package \cite{Mumax}.  
The microwave field takes the form
\begin{equation}
\mathbf{H}_{\text{mw}}=H_{\text{mw}}\left[\cos\phi(t)\hat{\mathbf{x}}+\sin\phi(t)\hat{\mathbf{y}}\right]\label{app}
\end{equation}
where \(H_{\text{mw}}\) is the amplitude of the microwave field and \(\phi(t)\)
is the phase. We consider a linear chirp whose instantaneous frequency
$f(t)\equiv \frac{1}{2\pi} \frac{\mathrm{d}\phi}{\mathrm{d}t}$ is linearly decreasing
with time for the changing rate $\eta$ (with unit of s$^{-2}$) as shown in Fig. 1(b),
\begin{equation}
f(t)=f_0-\eta t;\quad \phi(t)=2\pi (f_0t-\frac{\eta}{2}t^2),
\end{equation}
where $f_0$ is the initial frequency at $t=0$. The duration of the microwave pulse
is $T=\frac{2f_0}{\eta}$ so that the final frequency is $-f_0$.

\begin{figure}
	\includegraphics{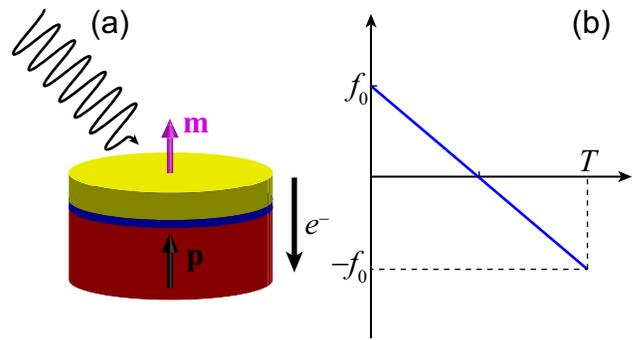}
\caption{\label{fig1} (a) Schematic diagram of the system. \(\mathbf{m}\) and \(\mathbf{p}\)
represent unit vectors of magnetization of free  and fixed layers respectively. 
A microwave field is applied onto the free layer, and a electric current flows through the 
spin valve. (b) The frequency of a down-chirp microwave (sweeping from \(+f_0\) to  \(-f_0\)). }
\end{figure}

\section{Numerical Results}

We first investigate the possibility to reverse the magnetization by a
down-chirp microwave pulse (DCMWP). At $t=0$, $m_z=1$ and the resonant 
frequency of the magnetization is $\gamma H_\text{K}=21.0$ GHz. 
Thus, to make the chirp microwave match the precession of $\mathbf{m}$ 
as much as possible, we use $f_0=\gamma H_\text{K}=21.0$ GHz. 
Fig. 2(a) shows the time evolution of $m_z$ under three different 
microwave fields. The red dashed line depicts the reversal by a down-chirp pulse of $f_0=21.0$ 
GHz, $\eta=67.2$ ns$^{-2}$ and $H_\text{mw}$ =0.045 T. The magnetization 
reverses fast with a switching time of 0.6 ns (throughout this paper, 
the switching time $t_\text{s}$ is defined as the time $m_z$ reaches $-0.9$).
As a comparison, the evolution of $m_z$ driven by a microwave of constant 
frequency (CFMW) 21.0 GHz and same amplitude $0.045$ T is shown by the black 
dash-dotted line. The magnetization only precesses around the initial state and 
does not reverse. To reverse the magnetization by a microwave of constant 
frequency within the same time (0.6 ns), the amplitude of the field has to be as 
large as 0.98 T as shown by the blue solid line, which is unrealistic in practice.
Therefore, DCMWP of small amplitude can induce subnanosecond magnetization 
reversal, showing significant advantage in comparison with conventional 
constant-frequency microwave driven schemes \cite{XR2006,T2016}. 
We then investigate how the switching time depends on the chirp rate 
$\eta$ and the microwave field amplitude $H_\text{mw}$.
According to the physical picture discussed in Section II, because the 
changing rate of the frequency should match the magnetization reversal, 
the switching time should be closed to the duration of the pulse.
Fig. 2(b) shows the $\eta$-dependence of the switching time $t_\text{s}$
for different $H_\text{mw}$. The length of the pulse is plotted with
green solid line for comparison. For each $H_\text{mw}$, there exists a
finite $\eta$ window in which the magnetization reversal occurs. 
Inside the window, the reversal time depends on $\eta$ non-monotonically 
due to the highly nonlinear magnetization reversal process. 
However, the reversal times oscillate near the length of the pulse, as 
shown by the green solid line, which justifies our physical picture. 
One can also see away from the critical values, the reversal time depends 
on $\eta$ and $H_\text{mw}$ weakly. That means a great flexibility in the 
choice of $\eta$ and $H_\text{mw}$ as well as the initial frequency, 
an extra nice property for applications. 
With $\eta=63.0$ ns$^{-2}$ and $H_\text{mw}=0.045$ T, the initial frequency
can be chosen in a wide range from 20.5 GHz to 39 GHz, with corresponding 
reversal time varying from 0.6 ns to 2 ns.
   \begin{figure*}[!t]
	\begin{center}
		\includegraphics[width=17cm]{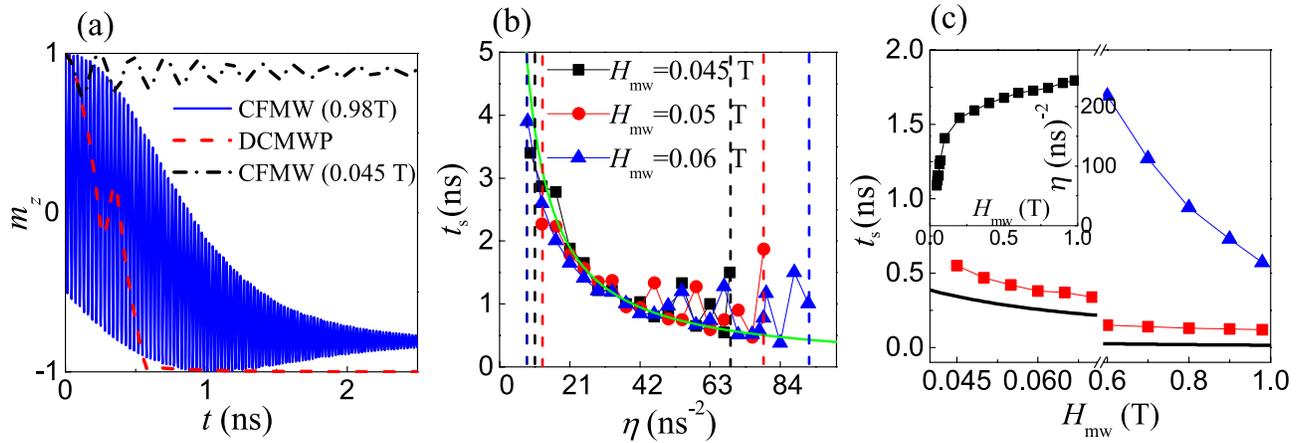}
	\end{center}
	
\caption{\label{fig2} (a) The time evolution of $m_z$ driven by
different sources: red dashed line for down-chirp microwave pulse (DCMWP)
of $f_0=21$ GHz, $H_\mathrm{mw}=0.045$ T and $\eta=67.2$ ns$^{-2}$, blue solid		
line for constant-frequency microwave of amplitude (CFMW) 0.98 T, black
dash-dotted line for CFMW of amplitude 0.045 T. (b) The dependence of switching 
times $t_s$	on the chirp rate $\eta$ for different microwave field amplitude
$H_\mathrm{mw}$. The vertical dashed lines are lower and upper limits of
$\eta$ for magnetization switching. (c) Comparison of magnetization
reversal times for different strategies. The horizontal axis is the field amplitude. 
Black solid line is the theoretical limit. Red squares/blue triangles are for the 
DCMWP/CFMW. Inset: the optimal chirp rates $\eta$ for different field amplitudes $H_\text{mw}$.}
\end{figure*}

To have a better sense of how good our strategy is, we compare the optimal 
reversal time of DCMWP of $f_0=21$ GHz and
$H_\text{mw}=0.045\sim0.92$ T (red squares) with the theoretical limit \cite{XR2007}
of the same field amplitude (black solid line) in Fig. 2(c). The corresponding
chirp rates for fastest reversal is shown in the inset. The reversal time
of CFMW of $f=21$ GHz is also shown (blue triangles). Below 0.6 T, only DCMWP
can switch the magnetization, with a subnanosecond reversal time that is only
a little longer than the theoretical limit. For field amplitude larger than 0.6 T,
the constant-frequency microwave is also able to switch the magnetization, but
the reversal time is much longer.

In order to have a better physical understanding of the fast switching
under DCMWP, we look at the magnetization process in more detail.
The red solid line in Fig. 3(a) shows the magnetization reversal process
driven by a down-chirp pulse of $f_0=21$ GHz, $H_\mathrm{mw}=0.045$ T and
$\eta=67.2$ ns$^{-2}$ [which is the same as the parameters
used in Fig. 1(a)]. Fig. 3(c) shows the trajectory of magnetization reversal.
Before (after) passing the equator, the magnetization rotates in
counterclockwise (clockwise) direction, as we discussed before. As a
comparison, we turn off the field at the moment that $\mathbf{m}$ just
passes the equator, so that afterwards the energy is purely dissipated by
Gilbert damping, i.e., the first term in the right-hand side of Eq. (3).
The black line in Fig. 3(a) shows the magnetization reversal
when the chirp field is turned off when $m_z=-0.004$. It is cleared that the reversal
process from the equator to $m_z=-1$ is much slower. Fig. 3(b) shows the
corresponding trajectory. Obviously, below the equator, the precessional
motion dominates and the longitudinal motion is slower than that in
Fig. 3(c). To further justify the physical picture, the down-chirp pulse
acts as an energy source (sink) before (after) crossing over the equator, we
observe the relative angle between the in-plane components of the
magnetization and the microwave field. From Eq. (3), the energy changing
rate due to the external field is
 \begin{equation}
I=-\mathbf{m}\cdot\dot{\mathbf{H}}_{\text{mw}}
=-H_{\text{mw}}\omega (t) \sin \theta(t) \sin \Phi(t),
\label{zeeman}
\end{equation}
where $\Phi(t)$ is the angle from $\mathbf{m}_t$ (the in-plane
component of $\mathbf{m}$) to $\mathbf{H}_\mathrm{mw}$.
We plot $\Phi(t)$ by blue line in Fig. 3(d), and $I$ by blue line in Fig.
3(e). Before $t=0.25$ ns, the
magnetization reverses quickly from $m_z=1$ to the equator as shown by
the red line. At the same time, $\Phi$ is between $0\sim -180^\circ$
around $-90^\circ$. Because the magnetization precesses counterclockwise
($\omega >0$),
this means $\mathbf{H}_\mathrm{mw}$ is $0\sim 180^\circ$ behind
$\mathbf{m}_t$. $I$ is positive so that the
microwave provides energy to the magnetization. When the $\Phi$ is
$-90^\circ$, the energy absorption rate reaches the maximum. Also,
in Fig. 3(e), the energy changing rate $I$ is positive.
Between 0.25 ns to 0.35 ns, the magnetization oscillates near the
equator because of the complicated nonlinear dynamics. After 0.35 ns,
the magnetization reverses from the equator to $m_z=-1$. At the same
time, $\Phi$ is again
between $0\sim -180^\circ$ around $-90^\circ$. But at this stage
the magnetization precesses clockwise ($\omega<0$). $\mathbf{H}_\mathrm{mw}$ is
$0\sim-180^\circ$ in front of $\mathbf{m}_t$.
$I$ is negative so that the
microwave absorbs energy from the magnetization. Also, in Fig. 3(e),
the energy changing rate $I$ is negative. Thus, the physical
picture of fast magnetization reversal by a down-chirp microwave pulse
is confirmed: for proper chirp rate and initial frequency, the
down-chirp microwave field matches the magnetization precession
in a large portion of the reversal process, so that before passing
the equator (the energy barrier) the microwave field provides energy
to the magnetization and after passing the energy barrier the
microwave field absorbs energy from the magnetization with a large
energy changing rate.

\begin{figure}[t!]
\begin{center}
\includegraphics[width=8.5cm, ]{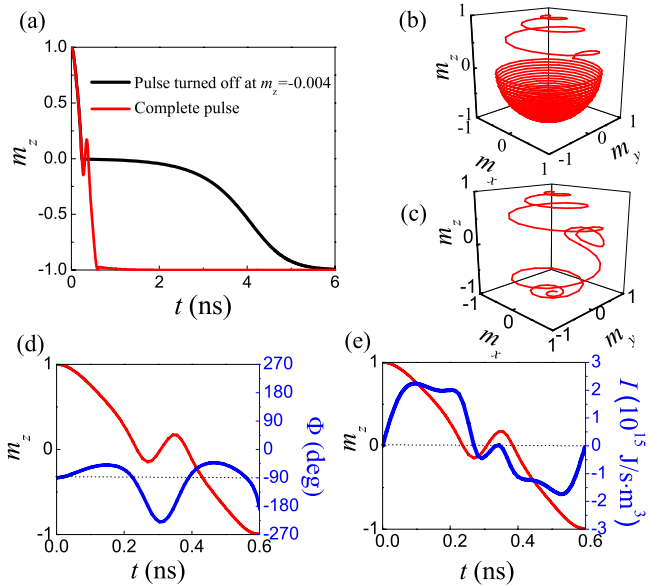}
\end{center}
\caption{\label{fig3} (a)  Magnetization reversal driven by a
down-chirp microwave field pulse of $f_0=21$ GHz, $H_\mathrm{mw}=0.045$ T and $\eta=67.2$ ns$^{-2}$. The red line is 
for a complete pulse. Black line shows the magnetization reversal 
if the pulse is turned off at \(m_{z}\)=-0.004. 
(b) Magnetization trajectory if the field is turned off when \(m_{z}\) = -0.004. (c) Trajectory of the magnetization for a complete pulse. (d) Plot of the relative angle $\Phi$ against time (blue line) and 
the time dependence of $m_z$ (red line). (e) Plot of the energy 
changing rate \(I\) of the magnetization against time (blue line) 
and the time dependence of $m_z$ (red line).}
\end{figure}

In the above studies, we used circular-polarized (CP) microwaves. 
Many microwave-generation methods, for example, the coplanar 
waveguide, generate linear-polarized (LP) microwaves. 
A LP microwave can be decomposed into the linear combination of two 
CP microwaves with opposite polarization. So a down-chirp LP 
microwave should also be capable to switch a magnetization particle. 
We numerically demonstrate this capability in Fig. 4(a)(b).
Figure 4(a) shows the chirp rate ($\eta$) dependence of switching 
time for a LP microwave of $H_\text{mw}=0.06$ T and $f_0=20$ GHz. 
The nanosecond magnetization reversal can be achieved in the window 
of $\eta=3.0\sim20$ ns$^{-2}$. Because of the other CP component, 
the magnetization dynamics becomes more complicated, as shown in 
Fig. 4(b), which plots the time evolution of $m_z$ for the optimal 
$\eta=20$ ns$^{-2}$. The complicated magnetization dynamics also 
results in different optimal initial frequency and chirp rate 
compared to the CP case. The optimal chirp rate is now $\eta=20$ 
ns$^{-2}$ for LP pulse which is smaller than the CP case, so that the 
switching time of LP pulse (2 ns) is also longer than that of CP pulse.

The obtained microwave magnetic field 0.045/0.06 T for CP/LP DCMWP is still too high. 
To further reduce its value, we can simultaneously apply a dc current. 
An electric current is polarized by the fixed layer so that it has a 
finite polarization along $z$-direction. Figures 4(c) and (d) show the 
$H_\text{mw}$-$J$ phase diagrams of the magnetization reversal for
CP and LP chirp microwave pulses respectively together with a dc current $J$. 
Below (above) the phase boundaries (shown by the blue lines), the 
switching time is longer (shorter) than 10 ns. The chirp pulses are 
chosen to be the ones that achieve fast reversal obtained before, i.e., $f_0=21$ GHz, $\eta=67.2$ ns$^{-2}$ for CP microwave and 
$f_0=20$ GHz, $\eta=20$ ns$^{-2}$ for LP microwave. 
If we require the switching time no longer than 10 ns, for 
magnetization reversal by electric current only, the required 
current density is about $1.4\times10^7$ A/cm$^2$; for magnetization 
reversal by CP/LP down-chirp microwave only, 
the minimal field amplitude is about 0.0445 T/0.06 T. 
Naturally, in the presence of both chirp wave and electric current, 
both $H_\text{mw}$ and $J$ can be smaller than the values above, 
which provides a large room to design practical magnetization 
reversal strategies according to the technical details.

 \begin{figure}[!t]
\includegraphics[width=8.5cm]{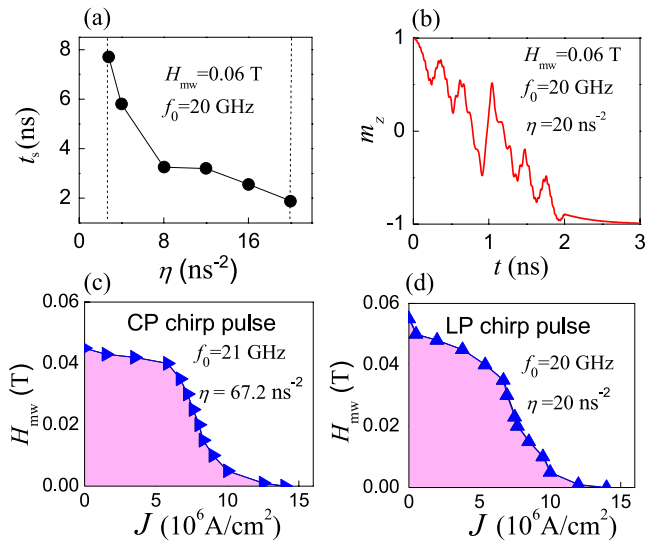}
\caption{\label{fig5} (a) Dependence of reversal time $t_\text{s}$ 
on the chirp rate for LP DCMWP of $H_\text{mw}=0.06$ T, $f_0=20$ GHz. 
(b) Time evolution of $m_z$ driven by LP DCMWP of $\eta=20$ ns$^{-2}$,
$H_\text{mw}=0.06$ T, $f_0=20$ GHz. 
(c-d) Phase diagram of magnetization reversal in terms of (c) CP and (d) LP DCMWP amplitude $H_\text{mw}$ and current density $J$. 
Pink region means the magnetization does not reverse or the
reversal time is longer than 10 ns. The white region means the magnetization reverses within 10 ns.}
\end{figure}

\section{Discussion and Conclusion}
Of course there are some previous studies of magnetization reversal by chirp microwave pulse\cite{KR,ZM,Suto17,GK}. But those pulses only provide energy to magnetization just to cross over the equator. After crossing over the equator, the magnetization goes to the reverse state slowly by dissipating the energy due to Gilbert damping. However, as mentioned before, we apply the DCMWP ( frequency profile of the pulse is shown in Fig. 1(b)) such that it acts as energy source (sink) for the magnetization  before (after) crossing over the equator which leads fast reversal.
    
The most challenging part of the DCMWP-driven magnetization reversal
is the generation of DCMWP with a wide band width and large chirp rates.
There are already several possible techniques for chirp-microwaves generation. 
In microwave photonics, there are several technologies of generating 
chirp microwaves \cite{Yao,Sanjeev}. Recently, it is found CP microwaves 
of time dependent frequency can be originated by coupling a magnetic 
nanoparticle to a pair of weak superconducting links \cite{Lcai,Lcai10}. 
The time dependency of the microwave frequency can be controlled by voltage. 
Another way to generate DCMWP is to use spin torque oscillator incorporating 
a field generating layer. By flowing time varying spin-polarized current 
through field generating layer, magnetization oscillation is excited.
 The oscillating magnetic moment in turn induces microwaves of time dependent frequency \cite{IEEE, IEEE2}. Therefore, the spin torque oscillator acts as source of microwave pulse, with the advantage that it is easy to be integrated with the spin valve to achieve good locality and scalability. There is already experimental realization to generate microwave 
of time dependent frequency \cite{ExSTO}. The widely-used coplanar waveguide
can also be used to generate DCMWP. Using two coplanar waveguides one can generate circular polarized DCMWP \cite{Suto17} while one coplanar waveguide can generate linear polarized DCMWP \cite{YNOZAKI}. Our findings provide improvements for fast magnetization reversal technologies with a clear physical picture, and shine a light on the future development of magnetic data storage and processing devices. 

In conclusion, we find a down-chirp microwave pulse can effectively reverse a
magnetic nanoparticle. Different from the magnetization reversal driven by 
the constant-frequency microwaves through synchronization that requires a 
strong field, the down-chirp microwave provides (absorbs) energy to (from) the 
magnetization before (after) the energy barrier, so that the reversal can be 
fast with a low field when the initial frequency and chirp rate are proper.
The down-chirp microwave pulse can be used together with a polarized 
electric current to design more practical reversal strategies.

\section{ACKNOWLEDGEMENTS}
This work was supported by the National Natural Science Foundation
of China (Grant No. 11774296) as well as Hong Kong RGC Grants No.
16300117 and No. 16301816. X. S. Wang acknowledges support from UESTC
and China Postdoctoral Science Foundation (Grant No.200 2017M612932). 
M. T. Islam acknowledges the Hong Kong PhD Fellowship.



\begin{thebibliography}{99}
 	 \bibitem{S2000} Shouheng Sun, C. B. Murray, D. Weller, L. Folks, and A.	Moser, Science \textbf{287}, 1989 (2000).


 \bibitem{SI2001} S. I. Woods, J. R. Kirtley, Shouheng Sun, and R. H.
 Koch, Phys. Rev. Lett.\textbf{ 87}, 137205 (2001).

 \bibitem{D2002}D. Zitoun, M. Respaud, M.-C. Fromen, M. J. Casanove,
 P. Lecante, C. Amiens, and B. Chaudret, Phys. Rev. Lett.\textbf{ 89}, 037203 (2002).

 \bibitem{B2001} B. Hillebrands and K. Ounadjela, eds., \textit{Spin Dynamics in Confined Magnetic Structures I \& II}, (Springer-Verlag, Berlin, 2001).

 \bibitem{nature} S. Mangin, D. Ravelosona, J. A. Katine1, M. J. Carey, B. D. Terris1 and Eric E. Fullerton. Nature Materials \textbf{5}, 210-215 (2006).


 \bibitem{1book} A. Hubert and R. Schafer, \textit{Magnetic Domains} (Springer, Berlin, 1998), Chap. 3.	

 \bibitem{XR2005} Z. Z. Sun and X. R. Wang, Phys. Rev. B \textbf{71}, 174430 (2005).

 \bibitem{STT}J. C. Slonczewski, J. Magn. Magn. Mater. \textbf{159}, L1 (1996); L. Berger, Phys. Rev. B \textbf{54}, 9353 (1996).

 \bibitem{M1998} M. Tsoi, A. G. M. Jansen, J. Bass, W.-C. Chiang, M. Seck,V. Tsoi, and P. Wyder, Phys. Rev. Lett. \textbf{80}, 4281 (1998); Y. B. Bazaliy, B. A. Jones, and S. C. Zhang, Phys. Rev. B \textbf{57}, R3213 (1998);  \textbf{69}, 094421 (2004).



 \bibitem{J1999} J. Sun, J. Magn. Magn. Mater. \textbf{202}, 157 (1999); Nature (London) \textbf{425}, 359 (2003).

 \bibitem{Z2004} J. Z. Sun, Phys. Rev. B  \textbf{62}, 570 (2000); Z. Li and S. Zhang, \textit{ibid.} \textbf{68}, 024404 (2003); \textbf{69}, 134416 (2004);   W. Wetzels, G. E. W. Bauer, and O. N. Jouravlev, Phys. Rev. Lett. \textbf{96}, 127203 (2006).






 \bibitem{A2000} A. Brataas, Y. V. Nazarov, and G. E. W. Bauer, Phys. Rev. Lett. \textbf{84}, 2481 (2000).

 \bibitem{X2000} X. Waintal, E. B. Myers, P. W. Brouwer, and D. C. Ralph, Phys. Rev. B \textbf{62}, 12 317 (2000); M. D. Stiles and A. Zangwill, \textit{ibid.}  \textbf{66}, 014407 (2002);



 \bibitem{E1999} E. B. Myers, D. C. Ralph, J. A. Katine, R. N. Louie, and R. A. Buhrman, Science \textbf{285}, 867 (1999).

 \bibitem{JA2000} J. A. Katine, F. J. Albert, R. A. Buhrman, E. B. Myers, and D. C. Ralph, Phys. Rev. Lett. \textbf{84}, 3149 (2000).

 \bibitem{R2004} R. H. Koch, J. A. Katine, and J. Z. Sun, Phys. Rev. Lett. \textbf{92}, 088302 (2004).





 \bibitem{HMORISE} H. Morise and S. Nakamura, Phys. Rev. B \textbf{71}, 014439 (2005); T. Taniguchi and H. Imamura,  \textit{ibid.} \textbf{78}, 224421 (2008).


 \bibitem{Tshinjo} Edited by T. Shinjo, \textit{Nanomagnetism and Spintronics} (Elsevier,	Amsterdam, 2009), Chap. 3.

 \bibitem{Joule} J. Grollier, V. Cros, H. Jaffr`es, A. Hamzic, J. M. George, G. Faini, J. B. Youssef, H. LeGall, and A. Fert, Phys. Rev. B \textbf{67}, 174402 (2003).



 \bibitem{XR2007} Z. Z. Sun and X. R. Wang, Phys. Rev. Lett. \textbf{97}, 077205 (2006); \textbf{98}, 077201 (2007).


 \bibitem{Bertotti} G. Bertotti, C. Serpico, and I. D. Mayergoyz, Phys. Rev. Lett. \textbf{86}, 724 (2001).


 \bibitem{IEEE} J.-G. Zhu and Y. Wang, IEEE Trans. Magn. \textbf{46}, (2010).

 \bibitem{XR2006} Z. Z. Sun and X. R. Wang, Phys. Rev. B  \textbf{73}, 092416 (2006);\textbf{74}, 132401 (2006).


 \bibitem{T2016} T. Taniguchi, D. Saida, Y. Nakatani, and H. Kubota,  Phys. Rev. B {\textbf 93}, 014430 (2016).

 \bibitem{C2003} C. Thirion, W. Wernsdorfer, and D. Mailly, Nat. Mater. \textbf{2}, 524 (2003).

 \bibitem{SI2006} S. I. Denisov, T. V. Lyutyy, P. H¨anggi, and K. N. Trohidou, Phys. Rev. B \textbf{74}, 104406 (2006).

 \bibitem{SO2008} S. Okamoto, N. Kikuchi, and O. Kitakami, Appl. Phys.
 Lett. \textbf{93}, 102506 (2008).



 \bibitem{TT2013} T. Tanaka, Y. Otsuka, Y. Furumoto, K. Matsuyama, and Y. Nozaki, J. Appl. Phys. \textbf{113}, 143908 (2013).




 \bibitem{KR} K. Rivkin and J. B. Ketterson, Appl. Phys. Lett. \textbf{89}, 252507 (2006).
 
  \bibitem{ZM} Z. Wang and M. Wu, J. Appl. Phys. \textbf{105}, 093903 (2009).

 \bibitem{Suto17}H. Suto, T. Kanao, T. Nagasawa, K. Mizushima, and R. Sato, Appl. Phys. Lett. \textbf{110}, 262403 (2017).
 
 \bibitem{GK} G. Klughertz, L. Friedland, P.-A. Hervieux, and G. Manfredi, Phys. Rev. B \textbf{91}, 104433 (2015).
 
 \bibitem{NBarros} N. Barros, M. Rassam, H. Jirari, and H. Kachkachi, Phys. Rev. B \textbf{ 83}, 144418 (2011); \textbf{88}, 014421 (2013).


 \bibitem{Lcai} L. Cai, D. A. Garanin, and E. M. Chudnovsky, Phys. Rev. B \textbf{87}, 024418 (2013).

 


 \bibitem{Topical view} S. Okamoto1, N. Kikuchi, M. Furuta, O. Kitakami and T. Shimatsu,   J. Phys. D: Appl. Phys. \textbf{48}, 353001 (2015).

 \bibitem{Yzhang} arXiv:1802.02415.

 \bibitem{Mumax} A. Vansteenkiste, J. Leliaert, M. Dvornik, M. Helsen, F. Garcia-Sanchez, and B. Van Waeyenberge, AIP Advances \textbf{4}, 107133 (2014).


\bibitem{Yao}W. Li and J. Yao, J. Lightwave Technol. \textbf{32}, 3573 (2014).

\bibitem{Sanjeev}S. K. Raghuwanshi, N. K. Srivastava, and A. Srivastava, Int. J. Electron.
\textbf{104}, 1689 (2017).

 \bibitem{Lcai10} L. Cai, and E. M. Chudnovsky,   Phys. Rev. B {\textbf 82}, 104429 (2010).

 \bibitem{IEEE2} J.-G. Zhu, X. Zhu, and Y. Tang, IEEE Trans. Magn. \textbf{44}, 125 (2008).

 \bibitem{ExSTO} H. Suto, T. Nagasawa, K. Kudo, K. Mizushima, and R. Sato, Nanotechnology \textbf{25}, 245501 (2014).



 \bibitem{YNOZAKI} Y. Nozaki, M. Ohta, S. Taharazako, K. Tateishi, S. Yoshimura, and K. Matsuyama, Appl. Phys. Lett. \textbf{91}, 082510 (2007).


   	









  \end{thebibliography}
\end{document}